\documentclass{iopart}

\usepackage{iopams}
\usepackage{epsfig}

\def\cm{\checkmark}

\begin{document}

\jl{4}

\title{Issues on NLO pQCD Programs}
\author{Thomas Hadig}
\address{H1 Collaboration\\
I. Physikalisches Institut, RWTH Aachen \\
D-52056 Aachen, Germany}

\begin{abstract}
This paper summarizes a talk presented at the Durham HERA '98 Workshop.
We compare the features that are available in NLO pQCD programs and
demonstrate that understanding where in phase space NLO calculations are
accurate is vital for extracting information from measurements at the
HERA experiments.
\end{abstract}


\noindent
A major task of the HERA experiments is to test pQCD features and the
extraction of the proton structure functions\cite{f2}. In the process of
comparing corrected data to theory, the calculation of quantities in
next-to-leading order is needed\cite{alphas, h1dijetgluon}.

Several programs allow predictions up to next-to-leading order for a
large set of variables. These programs and their features are compared
in section~\ref{features}. In addition to restrictions given by those
programs, it is also important to check the applicability of NLO pQCD
calculations for specific observables and the selected phase space.
Besides well known properties, like infrared safety\cite{mike98} or
factorizability, it has become clear lately that special care is needed
when using jet cuts\cite{h1lowQ}. One interesting topic in this region
will be pointed out in section~\ref{pt}.

\section{Features of NLO pQCD Programs}
\label{features}

During the last few years four multi-purpose pQCD calculation programs
for the HERA DIS processes have become available, Mepjet\cite{mepjet},
Disent\cite{disent}, Disaster++\cite{disaster}, and JETVIP\cite{jetvip}.
Their main features are given in table~\ref{tabfeature}. The most
important difference is the method used to handle cancelations of
singularities in real and virtual corrections. On the one hand, the
phase space slicing method integrates analytically in regions of
invariant masses lower than an extremly small cut-off parameter
$s_{\min}$\cite{psslic}. On the other hand, the subtraction method uses
the ``plus" prescription to calculate a counter term, that is subtracted
from the divergent distributions\cite{subtrac}. For several features,
e.g.~mass treatment and the contribution of resolved and electroweak
processes, only one program is available; therefore cross checks of
results, where these corrections get important, are impossible.

\begin{table}[b]
\begin{indented}
\item[]
\fl\begin{tabular}{@{}lllll}
\br
&MEPJET &DISENT &DISASTER++ &JETVIP \\
\mr
version & 2.2 & 0.1 & 1.0.1 & 1.1\\
\mr
method & PS slicing & subtraction & subtraction & PS slicing \\
\mr
1+1,2+1 & NLO     & NLO    & NLO    & NLO \\
3+1     & LO      & LO     & LO     & LO \\
4+1     & LO      & ---    & ---    & --- \\
jet shapes & LO  & LO     & LO     & LO \\
\mr
full event record & \cm & \cm & \cm & (\cm) \\
\mr
scales       & all    & factorisation: $Q^2,$ fixed, & all & all \\
             &        & renormalisation: all &  & \\
\mr
flavour dependence & switch & switch & full & switch \\
\mr
quark masses & LO     & ---    & ---  & --- \\
resolved contribution  & --- & --- & --- & NLO \\
electroweak contribution & LO & --- & --- & --- \\
polarized x-section  & NLO & --- & --- & --- \\
\br
\end{tabular}
\end{indented}
\caption{Comparison of the different features of NLO pQCD programs.}
\label{tabfeature}
\end{table}

\section{Comment on DiJet cuts}
\label{pt}

Using dijet measurements several QCD tests and parameter extractions
have been performed by the HERA experiments\cite{alphas,h1dijetgluon}.
NLO pQCD programs are important tools needed for this task, but the
measurements nowadays tend to enter regions where pQCD alone is not able
to describe the data, e.g.~in the transition region from photoproduction
to DIS, where resolved processes become important. In those regions it
is all the more important to have reliable NLO predictions.

An important restriction for cuts on dijet photoproduction calculations
was pointed out some time ago by Frixione and Ridolfi\cite{frixrido}.

\begin{figure}[bht]
\fl
\begin{center}
\epsfig{file=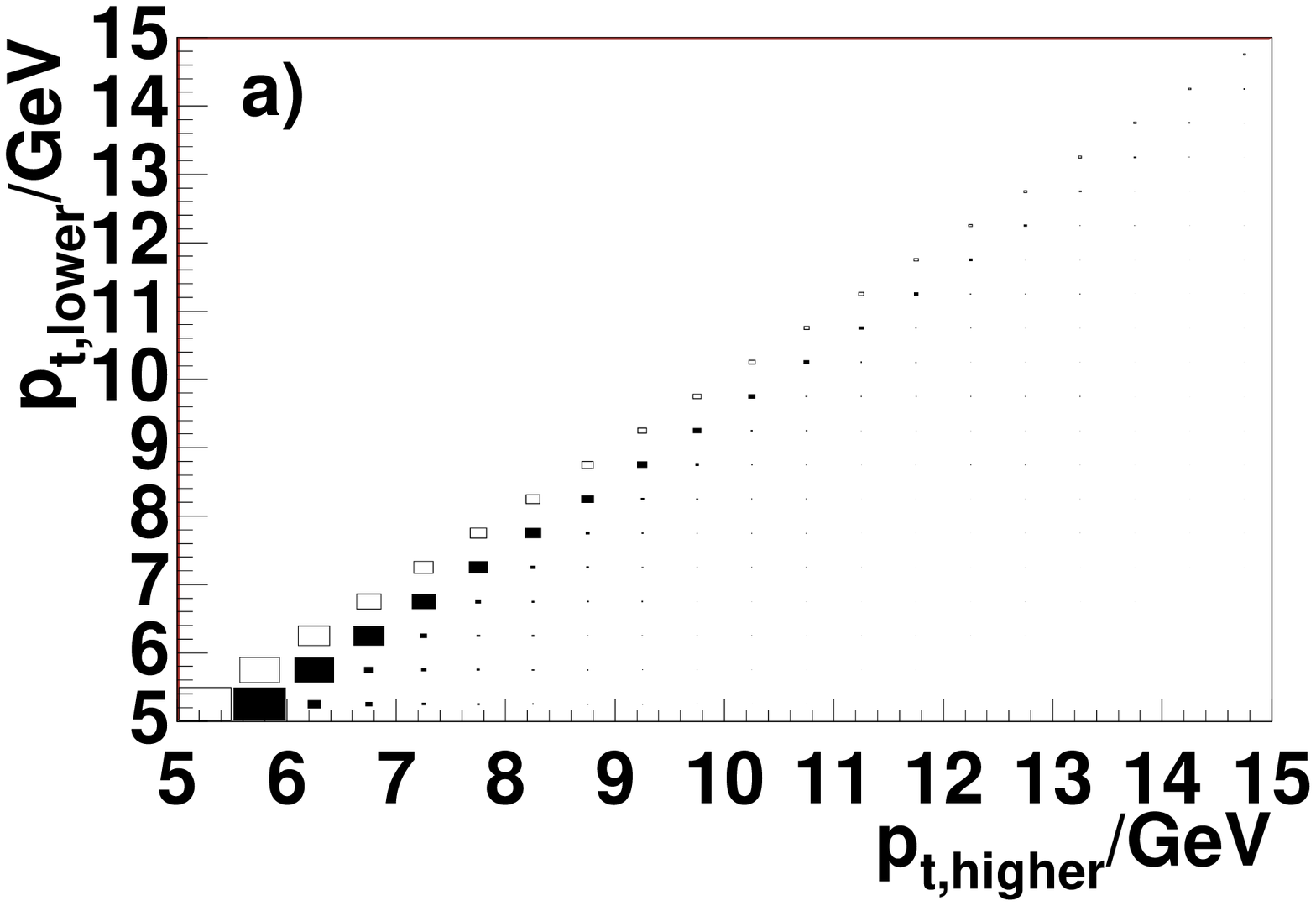,width=0.45\hsize}
\epsfig{file=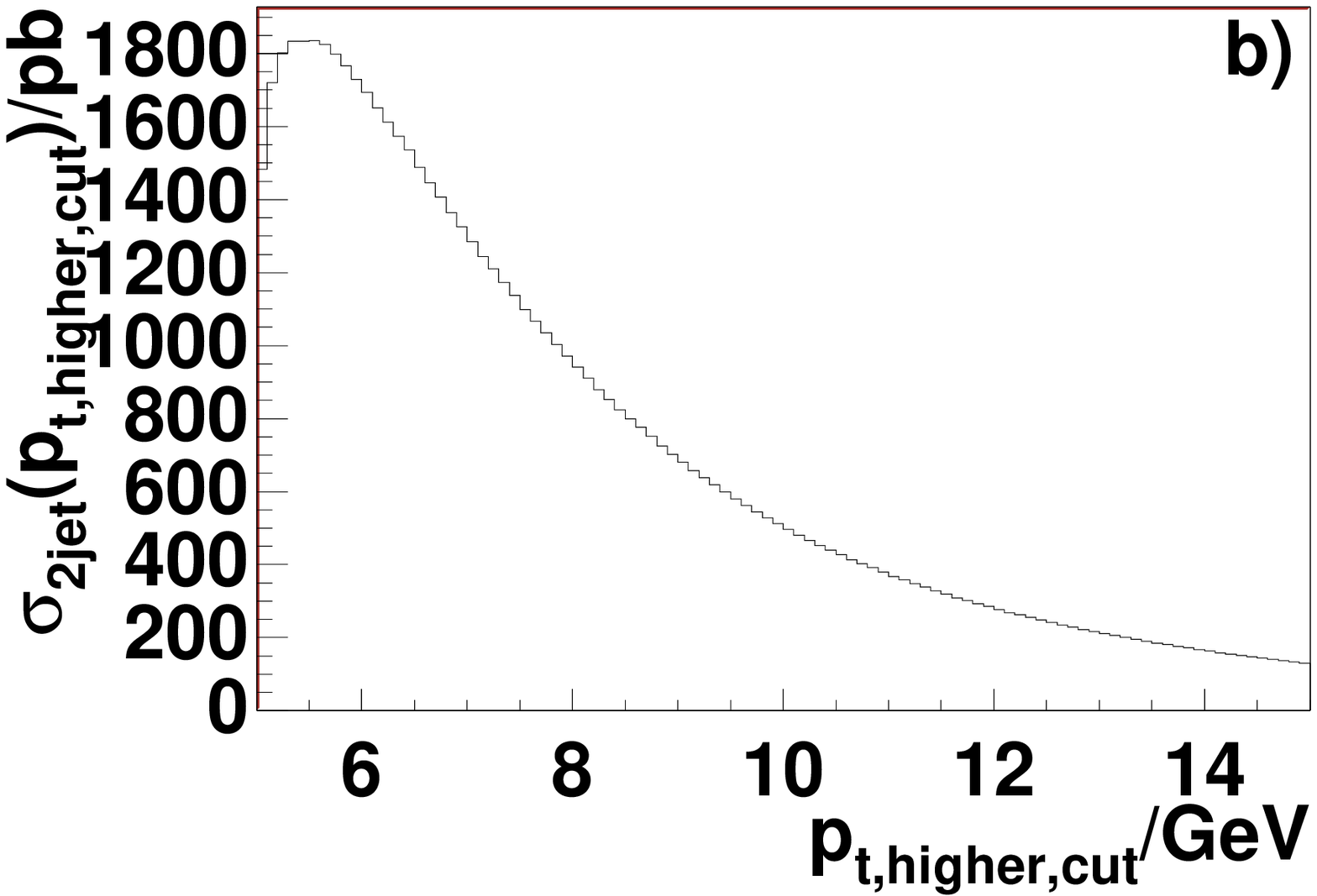,width=0.45\hsize}
\end{center}
\caption{a) Distribution in bins of the transverse jet momenta of each jet.
The size of the box corresponds to the dijet cross section in that bin.
Hollow boxes denote negative values, full positive. The calculation was
done using Disent. b) Integrated dijet cross section as a function of
the highest jet transverse momentum after imposing a cut of
$p_{t,\mbox{\scriptsize lower}}> 5\mbox{GeV}.$}
\label{figpt}
\end{figure}

The same arguments also hold for DIS processes. In figure~\ref{figpt}a) the NLO
prediction of the double differential cross section in transverse momenta of
both jets in the Breit frame is shown. The jets were found using the
longitudinal boost invariant $k_t$ algorithm\cite{ktclus}. For the LO
contributions the jet $p_t$ are balanced. The same is true for the virtual
corrections and this leads to large corrections on the diagonal, which are
known to be negative. The real corrections, canceling the virtual divergences,
introduce differences in the jet $p_t.$ After imposing a cut of
$p_{t,\mbox{\scriptsize lower}} > 5\mbox{GeV}$ for both jets,
figure~\ref{figpt}b) shows the integrated cross section as a function of the
highest jet $p_t.$ This plot clearly shows, that the NLO prediction breaks down
at a $p_{t,\mbox{\scriptsize higher}}$ cut of approximately $6\mbox{GeV},$ since
lowering the $p_t$ cuts and thus enlarging the allowed phase space leads to a
reduced NLO prediction.

A possible solution is to introduce an asymmetric cut, where the difference
of the cut values ensures, that the NLO prediction is on the falling
edge of the distribution of figure~\ref{figpt}b). A different Ansatz
is to make a symmetric cut and an additional cut on the sum of the
jet $p_t$ values, e.g.~$\sum_{1,2} p_{t,i} > 17\mbox{GeV}.$ This cut
removes the main negative contribution in the lower left corner of
figure~\ref{figpt}a) and ensures, that the cancelation of singularities
takes place\cite{h1lowQ}.


\section{Conclusions}

An overview of features currently available in NLO pQCD programs for
HERA deep inelastic scattering processes has been given. In addition a
special issue on jet cuts for dijet production has been pointed out
demonstrating, that asking for a symmetric minimal $p_t$ on both jets
leads to unreliable NLO predictions. Two alternative scenarios are
given, that produce reasonable results.


\ack

{\small
I would like to thank R.~Devenish, W.J.~Stirling, and M.R.~Whalley for
organising this interesting and fruitful workshop and J.~Butterworth,
M.H.~Seymour, and G.~Thompson for the kind invitation. Special thanks
goes to Ch.~Berger and S.J.~Maxfield for proofreading and commenting
on this contribution.}



\end{document}